\documentclass[12pt,titlepage]{article}
\usepackage{epsfig}
\usepackage{graphicx}
\usepackage{amsmath,amssymb}
\usepackage{bm}
\usepackage{color}
\usepackage{empheq}
\usepackage{slashed}
\usepackage{cite}
\usepackage[all]{xy}
\newcommand{\beq}{\begin{equation}}
\newcommand{\eeq}{\end{equation}}
\newcommand{\bea}{\begin{eqnarray}}
\newcommand{\eea}{\end{eqnarray}}
\newcommand{\bean}{\begin{eqnarray*}}
\newcommand{\eean}{\end{eqnarray*}}

\newcommand{\bepsilon}{{\bm \epsilon}}

\newcommand{\bdelta}{{\bm \delta}}
\newcommand{\hbdelta}{\widehat{\bm \delta}}
\newcommand{\WQ}{Q}
\newcommand{\WJ}{J}
\newcommand{\NJ}{{\cal J}}
\newcommand{\NQ}{{\cal Q}}

\begin{document}

\title{
\vskip -200pt
\flushright{\small DIAS-STP-18-05}\\
\vskip 200pt
\begin{center}
A tale of two derivatives: phase space symmetries and Noether charges in diffeomorphism invariant theories
\end{center}
}

\author{{Brian P. Dolan}\\
{\small Department of Theoretical Physics, National University
of Ireland,}\\
{\small  Maynooth, Ireland}\\
{\small\textit {and}}\\
{\small Dublin Institute for Advanced Studies,
10 Burlington Rd., Dublin, Ireland}\\
{\small e-mail: \texttt{bdolan@thphys.nuim.ie}}}

\maketitle

\begin{abstract}
For a field theory 
that is invariant under diffeomorphisms there is a subtle interplay between symmetries, conservation laws and the phase space of the theory.  The natural language for describing these ideas is that of differential forms and both differential forms on space-time and differential forms on the infinite dimensional space of solutions of the equations of motion of the field theory play an important role.  There are exterior derivatives on both spaces and together they weave a double differential complex which captures the cohomology of the theory. This is important in the definition of invariants in general relativity, such as mass and angular momentum
and is also relevant to the study of quantum anomalies in gauge theories.

We derive the structure of this double complex and show how it relates to 
conserved quantities in gravitational theories.  One consequence of the construction is that conserved quantities can be calculated exactly at finite
distance --- for example it is not necessary to go to asymptotic regimes to calculate the mass or angular momentum of a stationary solution of Einstein's equations, 
they can be obtained exactly by an integration over any sphere outside the mass even at finite radius.

\end{abstract}
\section{Introduction}

Conserved quantities are a consequence of Noether's theorem. The simplest version of Noether's theorem assumes that the
Lagrangian governing the dynamics is invariant under some symmetry operation.  
Under a general variation of the  fields $F^I$
the variation of the Lagrangian density $L(F)$, viewed as a differential form in an $(n+1)$-dimensional space-time ${\cal M}$, is
\beq \delta L(F^I) = E_J(F^I) \delta F^J + d \theta\label{eq:delta-L}\eeq
where $E_I(F^I)=0$ are the equations of motion and $\theta(F^I,\delta F^I)$
is an $n$-form which depends on the fields and their variation.  
When there is symmetry of the dynamics Noether's theorem gives an associated conserved charge.  The simplest case is when $L$ itself is an invariant of the
symmetry.  If $\delta_{\NQ}=\epsilon {\cal T}_{\NQ}$ and $\delta_{\NQ} L=0$, where ${\cal T}_{\NQ}$  is a 
symmetry generator with an associated charge $ {\NQ}$, then
\[\delta_{\NQ} L = d \theta(S^I,\delta_{\NQ} S^I)=\epsilon d \theta(S^I,{\cal T}_{\NQ} S^I)=0,\]
where $S^I$ is a solution of the equations of motion.
The usual Noether current associated with the conserved charge $ {\NQ}$ is the Hodge dual of $\theta(S^I,{\cal T}_{\NQ} S^I)$
\[\theta = *j\]
and it is common in field theory to express the conservation of the Noether
current as a zero divergence condition,
\[d*j =0.\]
  Integrating over a region of space-time ${\cal M}$ bounded by space-like hypersurfaces $\Sigma$ and $\Sigma'$ 
gives\footnote{If $\Sigma$ and $\Sigma'$ have boundaries the fields are assumed to vanish
sufficiently fast there that there is no contribution from any time-like
component of $\partial {\cal M}$.}
\[\int_{\cal M} d \theta = \int_\Sigma \theta - \int_{\Sigma'} \theta =0. \] 
Thus the charge
\beq  \NQ=\int_\Sigma * j = \int_{\Sigma'} *j\label{eq:Noether-current}\eeq
associated with the the symmetry and the chosen solution is an invariant --- independent of the space-like hypersurface
on which it is evaluated.  With Maxwell's equations, for example,
\[ d*F = 4 \pi *j\]
so $*j$ is exact when the equations of motion are satisfied
and the electric charge is obtained from Gauss' law,
\[ \NQ=\frac{1}{4\pi}\int_\Sigma d *F=\frac{1}{4\pi}\int_{\partial \Sigma} *F,\]
where $\partial \Sigma$ is the boundary of $\Sigma$, taken to envelop
all the charges.

 Under diffeomorphisms $L$ itself is not invariant, it changes by a surface term and it is the action that is invariant, not the Lagrangian density.  
Conserved currents can be defined using the energy-momentum tensor $T_{a b}$:
when $\vec K$ is a Killing vector $j^a = T^a{}_b K^b$ is a conserved current. 
This requires splitting $L$ into \lq\lq geometrical'' part, such as the Einstein Lagrangian, and a \lq\lq matter'' part from which $T_{a b}$ is derived.

 Things get more interesting when a classical symmetry is broken quantum mechanically and we have a quantum anomaly.  The Stora-Zumino descent equations then provide
a powerful tool for classifying and understanding anomalies \cite{Stora-Zumino}
and quantum anomalies can break diffeomorphism invariance, or equivalently local Lorentz invariance \cite{Gaume-Ginsparg}. 

The full story
goes much deeper however and is related to covariant symplectic 
structures on the phase space of the theory. It transpires that
$\theta$ is a 1-form on the co-tangent bundle $T^*{\cal S}$ of the space of solutions ${\cal S}$ and 
is related to a symplectic potential 
associated with the phase space of the dynamical theory.

In this work it will be shown how all of these these ideas fit into
the same mathematical structure of differential complexes and that the relevant 
conserved quantities are related to cohomology classes. 
Among the key ingredients are a general co-ordinate invariant action and a solution of the equations of motion with a Killing vector generating the symmetry.  
 
Some years ago Crnkovi\'c and Witten  \cite{C-W} gave a method for constructing a symplectic form on the space of solutions of the equations of motion of a relativistic field theory.  They used their formalism to obtain the appropriate symplectic forms for Yang-Mills theory and for general relativity in 4-dimensions.  Their construction provides a co-variant description of relativistic field theories in the phase space of solutions modulo gauge transformations and diffeomorhpisms
${\cal G}$, $\widehat {\cal S}={\cal S}/{\cal G}$,
which is ideally suited to studying symmetries and conserved quantities.
The idea of a symplectic structure for diffeomorphism invariant theories 
was first introduced in \cite{Friedman} to investigate
instabilities in rotating relativistic fluids. 
Wald and collaborators have generalized Crnkovi\'c and Witten's formalism to
a very wide class of diffeomorphism invariant theories in \cite{Lee+Wald,Wald1,Wald2, Iyer+Wald} and studied conserved quantities associated with Killing symmetries, such as angular momentum in rotationally invariant solutions and mass in stationary solutions.  

This formalism was shown in \cite{Iyer+Wald} to reproduce the ADM mass \cite{ADM} for stationary asymptotically flat black holes in Einstein gravity.  At the same time it clarifies the origin of the mysterious factor of two that is well known to arise when comparing the Komar mass with the ADM mass \cite{Komar,Katz}.

It is shown elsewhere \cite{WD} that the Brown-York mass \cite{Brown+York} 
also has a natural interpretation within the framework
of Lee and Wald's formalism \cite{Lee+Wald}.  The Brown-York mass is defined in terms of the difference of the extrinsic curvature of the sphere at infinity for a given solution and the extrinsic curvature of the sphere at infinity in flat space-time. Being the difference of two solutions of the equations of motion this is a 1-form on ${\cal S}$. It can also be shown \cite{WD} that Lee and Wald's formalism reproduces the Bondi mass \cite{Bondi} in stationary space-times.

The construction in \cite{Lee+Wald} is general enough to include theories with a cosmological constant $\Lambda$, of either sign when $\Sigma$ is compact without boundary. When $\Sigma$ has a boundary one restricts to negative $\Lambda$ so that
the asymptotic regime of a black hole solution is well defined. 
Lee and Wald's expression then agrees with the 
variation of the Henneaux and Teitelboim mass for asymptotically anti de-Sitter (AdS) Kerr black holes in 4-dimensions \cite{Henneaux-Teitelboim,GPP}.  
It is not immediately obvious that the Wald and Henneaux-Teitelboim masses are the same, the details of the calculation are non-trivial and are given elsewhere \cite{Hajian+Jabbari,Skenderis}.

\section{The symplectic form on the space of solutions\label{sec:Wald}}

In this section we review Wald's original construction of the 
symplectic structure on the space of solutions and the associated
Noether charge arising from diffeomorphism invariance and show how it is
described by a double complex structure.
The analysis is general enough to include gauge theories and general 
diffeomorphic invariant theories of gravity, both in metric formulations and in terms of local Lorentz frames, vielbeins and connections. 
 
Let ${\cal F}$ be the space of fields and ${\cal S}$ the space of solutions and denote the fields by $F^I\subset{\cal F}$, indexed by $I$.  For simplicity we shall assume that all the fields, including the metric, are dynamical --- the situation when the metric is non-dynamical is discussed in \cite{Iyer+Wald}.

A solution of the equations of motion will be
be denoted by $S^I\subset{\cal S}$.
We shall primarily be interested in field configurations which solve the
equations of motion and functionals will depend not only on the fields $S^I$ (primary fields) but also on their partial derivatives of order $k$, $\partial_k S^{I_k}$
(descendants), indexed by $I_k$.  

Consider an action which is an integral over an
$(n+1)$-dimensional manifold ${\cal M}$ with a Lagrangian density $L(F^I)$,
 \beq {\cal A}\big[F^I\big] = \int_{\cal M} L\big(F^I\big).\eeq 
Under a variation of the  fields the Lagrangian changes from
$L$ to $L + \delta L$ with
\beq  \delta L =   {E }_I \wedge \delta F^I  + d\theta\label{deltaL}\eeq
where
\beq {E}_I(F^I)=0\label{EoM}\eeq
are the equations of motion,
 $\theta(F^I,\delta F^I)$ is an $n$-form on ${\cal M}$.

Under a second variation of the  fields we define \cite{C-W, Lee+Wald, Wald2} 
\beq 
\omega(F^I,\delta_1 F^I,\delta_2 F^I):=\delta_1  \theta_2 - \delta_2\theta_1 \label{omegaTheta12}
\eeq
where $\theta_1=\theta(F^I,\delta_1 F^I)$ and $\theta_2=\theta(F^I,\delta_2 F^I)$.
Demanding that $F^I$ are a solution of the equations of motion determines a point in the space of solutions, ${\cal S}$.
We further demand that $\delta_1 F^I$ and $\delta_2 F^I$ are solutions of the linearised equations of motion (denoted $\delta_1 S^I$ and $\delta_2 S^I$) in the sense that
\beq{E }_I(S^J + \delta S^J) \approx {E }_I(S^J) + \delta {E }_I(S^J)= \delta {E }_I(S^J)= 0,
\label{lEoM}\eeq
to first order in $\delta E^J$.
Then $\theta$ is a 1-form on $T^*{\cal S}$ and $\omega$ a 2-form (they are both $n$-forms on $T^*{\cal M}$). 

In this context a field variation $\delta$ can be viewed as the exterior derivative on the space of solutions, with $\delta^2=0$.
We shall adopt the convention that boldface symbols represent differential forms
on ${\cal S}$ and write
\beq {\bm \omega} = {\bm \delta} {{\bm \theta}}.\label{eq:omegadeltatheta}\eeq

From now on it will always be assumed that $S^I$ satisfy the equations of motion and all ${\bm \delta} S^I$ satisfy the linearised equations of motion (\ref{lEoM}), 
so we are always dealing with differential forms on ${\cal S}$.
For brevity we shall refer to such field configurations as \lq\lq on-shell''.
Thus equation (\ref{deltaL}) is
\beq {\bm \delta} L= d {\bm \theta}\label{eq:deltaL}\eeq
on-shell and this then implies that
\beq 0={\bm \delta}^2 L = {\bm \delta} d  {\bm \theta}
= d {\bm \delta} {\bm \theta} = d {\bm \omega},\label{eq:domega0}\eeq
since the exterior derivative $d$ on ${\cal M}$ does not depend on the fields and $d{\bm \delta} = {\bm \delta} d$.
Hence ${\bm \omega}$ is $d$-closed on-shell.

Adding a total derivative to the action density does not change the equations of motion but can 
change ${\bm \omega}$ by a total derivative. If
\[ L \rightarrow L' = L  + d\alpha\]
then, again on-shell,
\[ {\bm \delta} L' = {\bm \delta} L  + {\bm \delta} d \alpha =  {\bm \delta} L
+ d {\bm \delta} \alpha = d ({\bm \theta} + {\bm \delta} \alpha)=\bdelta{\bm \theta}'\]
with
\beq{\bm \theta}' =  {\bm \theta} + {\bm \delta} \alpha + d{\bm \psi},
\label{eq:psi-def}\eeq
where ${\bm \psi}$ is an $(n-2)$-form on ${\cal M}$ and a 1-form on ${\cal S}$.
Now
\[ {\bm \delta} {\bm \theta}' =  {\bm \delta} {\bm \theta} + d {\bm \delta} {\bm \psi}\]
so
\[ {\bm \omega}'= {\bm \omega} + d {\bm \delta}{\bm \psi}.\] 
Thus if $\Sigma$ is a hypersurface in ${\cal M}$ which is compact without boundary
\beq
{\bm \Theta}':=\int_\Sigma {\bm \theta}' = \int_\Sigma {\bm \theta} + {\bm \delta} \left(\int_\Sigma \alpha\right)
= {\bm \Theta}+ {\bm \delta} \left(\int_\Sigma \alpha\right)
\label{pre-symplecticpotential}\eeq
and
\beq {\bm \Omega}=\bdelta {\bm \Theta}=\int_\Sigma {\bm \omega} = \int_\Sigma {\bm \omega}' = \bdelta {\bm \Theta}'\label{pre-symplecticform}\eeq
 is unchanged. If $\Sigma$ has a boundary $\partial \Sigma$ these expression are still valid
 provided  $\int_{\partial\Sigma}{\bm \psi}=0$.

On the other hand the variation of the action
\[ {\bm \delta}{\cal A} = \int_{\cal M} (d {\bm \theta} +  E_I \wedge {\bm \delta} S^I) =\int_{\cal M} d {\bm \theta} = \int_{\partial{\cal M}} {\bm \theta}\]
on-shell so, since ${\bm \delta}^2=0$,
\[{\bm \delta}^2 {\cal A}= \int_{\partial{\cal M}} {\bm \delta} {\bm \theta}
=  \int_{\partial{\cal M}} {\bm \omega}=0.\]

If $\partial {\cal M}$ consists of two space-like hypersurfaces $\Sigma$ and $\Sigma'$, connected
by a time-like tube,\footnote{\textit{i.e.} an $n$-dimensional space with one time-like direction 
and $(n-1)$ space-like directions.
} on which the  fields and ${\bm \theta}$ vanish then, with appropriate orientations for $\Sigma$ and $\Sigma'$,
\[   \int_{\partial{\cal M}} {\bm \omega} =  \int_\Sigma {\bm \omega}- \int_{\Sigma'} {\bm \omega}=0\]
and
 \beq {\bm \Omega}=\int_\Sigma {\bm \omega} 
= \int_{\Sigma'} {\bm \omega} = \int_{\Sigma} {\bm \omega}'\label{pre-symplecticformSigma}\eeq
is independent of the space-like hypersurface chosen and of any total derivatives added to the Lagrangian.

If $\Sigma$ is a Cauchy surface then ${\bm \Omega}$ is a pre-symplectic form on ${\cal S}$, in the sense of \cite{C-W}, and
${\bm \Theta}$ is a pre-symplectic potential. They are not quite symplectic 
because we still need to factor out diffeomorphisms and restrict ${\cal S}$ to $\widehat{\cal S}={\cal S}/{\cal G}$. 

The above discussion is summarised in the on-shell commutative diagram below:
\beq
\xymatrix@=50pt@M=10pt{
L+d\alpha \ar[r]_{\kern -10pt \bm \delta} & {{\bm \delta} L=d{\bm \theta}} \ar[r]_{\bm \delta} & 0 & \\
\alpha \ar[u]_d \ar[r]_{\kern -30pt \bdelta }& \ar[u]_d  {\bm \theta} + \bdelta \alpha + d{\bm \psi}  \ar[r]_{\bm \delta} & 
{\begin{array}{r}
\bdelta{\bm \theta} + \bdelta d{\bm \psi}  \\
=  {\bm \omega} + d\bdelta{\bm \psi}
\end{array}}
\ar[u]_d\ar[r]_{\kern 30pt \bm \delta} & 0\\
& {\bm \psi} \ar[u]_d \ar[r]_\bdelta &  \bdelta {\bm \psi} \ar[u]_d \ar[r]^\bdelta
& 0.} \label{eq:xy-omega-delta-theta} \eeq

Mathematically the structure here is that of a doubly graded complex \cite{Bott+Tu} and, as for any such complex,  it can easily be reduced to a single complex. 
Let $W^{p,q}$ denote the space of $p$-forms on ${\cal S}$ and $q$-forms on ${\cal M}$. Then a singly graded complex is obtained by taking the space of constant
total degree $r$, 
\[W^r=\bigoplus_{p+q=r}W^{p,q}\]
and defining a differential operator 
\[{\bm D}=\bdelta +(-1)^p d\]
acting on $W^r$ and sending $W^r \rightarrow W^{r+1}$, with ${\bm D}^2=0$.
Then
\[ \xymatrix@=50pt@M=10pt{
W^0 \ar[r]_{\bm D} &  W^1 \ar[r]_{\bm D} &  \cdots  \ar[r]_{\bm D} &  W^r \ar[r]_{\bm D} & \cdots
}\]
is a singly graded complex. 
In particular
\bean
    {\bm D} \big\{(L+d\alpha) + ({\bm \theta} + \bdelta \alpha  - d {\bm \psi}) 
    + \bdelta {\bm \psi} \big\}
    \kern -150pt \ &&\\
&=& (\bdelta L + \bdelta d\alpha) + (\bdelta {\bm \theta}  - \bdelta d {\bm \psi}
-d {\bm \theta} -d\bdelta \alpha) + d\bdelta {\bm \psi}\\
&=& \bdelta {\bm \theta} \ =\
  {\bm \omega}.
  \eean
 
Integrating the $n$-forms ${\bm \theta}$ and ${\bm \omega}$ over $\Sigma$ gives the following cohomology on the space of solutions
\[\xymatrix@=50pt@M=10pt{ {\bm \Theta} \ar[r]_{\bm \delta} & {\bm \Omega}  \ar[r]_{\bm \delta}
& 0.  }\] 

\section{Diffeomorphisms and Killing symmetries}

\subsection{General diffeomorphisms}

We would like to understand how diffeomorphisms fit into this picture.
Diffeomorphisms are generated by an infinitesimal vector field $\epsilon\vec X$ with $\epsilon$ an infinitesimally small constant. $\vec X$ generates a variation of the fields 
\beq {\bm \delta} F^I = \epsilon {{\cal L}}_{\vec X}\, F^I = \epsilon d i_{\vec X}\, F^I +  \epsilon i_{\vec X}\, d F^I\:.\eeq
One might expect that, since $\vec{X}$ is a fixed vector field independent of the fields $F^I$, 
\[ {\bm \delta} {{\cal L}}_{\vec X} = {{\cal L}}_{\vec X} \,{\bm \delta} \qquad \hbox{and} \qquad
{\bm \delta} {i}_{\vec X} = {i}_{\vec X}\, {\bm \delta}\]
but we should remember that we are interpreting ${\bm \epsilon{\cal L}}_{\vec X}\, F^I$ as a 1-form on ${\cal S}$.
It is convenient to promote $\epsilon$ to be a constant Grassmann parameter ${\bm \epsilon}$, 
where ${\bm\epsilon}$ is a 1-form on ${\cal S}$ which anti-commutes with $\bdelta$, and write
\beq \bdelta_{\vec X} ={\bm \epsilon}{\cal L}_{\vec X}\eeq
with
\[\bdelta_{\vec X}\, \bdelta =- \bdelta \bdelta_{\vec X}.\]
This maintains the condition $d{\bm \delta} = {\bm \delta} d$, since ${i}_{\vec X}\, {\bm \delta}=
{\bm \delta}{i}_{\vec X}\, $.

It is then natural to decompose the exterior derivative ${\bm \delta}$ 
on ${\cal S}$ into a genuine physical variation of the fields $\widehat{\bm \delta}$ and a variation arising from 
diffeomorphisms,
\beq \bdelta =\hbdelta + \bdelta_{{\vec X}}\, = \hbdelta + \bepsilon {\cal L}_{\vec X}\,.
\label{eq:deltahatdef}\eeq
If the solution depends on a set of parameters (moduli) then the variation $\hbdelta$ can be induced by varying the parameters --- it is an exterior derivative on the moduli space.  
This decomposition is unique, if there were different decompositions
\[ \bdelta = \hbdelta + \bdelta_{\vec X} = \hbdelta' + \bdelta_{\vec X\,{}'}\]
then
\[ \hbdelta - \hbdelta' = \bdelta_{\vec X} - \bdelta_{\vec X{}'}\] 
and both sides must vanish since the left-hand side is a genuine physical variation of the fields and the right-hand side
is a diffeomorphism.

There is a similar decomposition of forms, any 1-form ${\bm \eta}$ on ${\cal S}$ (and $q$-form on ${\cal M}$) can be decomposed as
\beq  {\bm \eta}=  \widehat{\bm \eta} + {\bm \eta}_{\vec X}
= \widehat{\bm \eta} + \bepsilon \eta(\vec X\,)
\label{eq:epsilon-split}\eeq
where ${\bm \eta}_{\vec X}=\bepsilon \eta(\vec X)$ and $\eta(\vec X\,)$ is a $q$-form on ${\cal M}$ and a function on ${\cal S}$.
In particular 
\bea {\bm \theta}= \widehat{\bm\theta} + {\bm\theta}_{\vec X}  = \widehat{\bm\theta} + \bepsilon \theta(\vec X\,) 
\eea
where $\theta(\vec X\,)$ is an $n$-form on ${\cal M}$. Then
\beq d{\bm \theta} =  \bdelta L 
\qquad \Rightarrow \qquad 
\left\{\begin{array}{ll}
d\widehat{\bm \theta} & = \ \hbdelta L,\\ 
d \theta(\vec X\,) & = d i_{\vec X} L.
\end{array}\right.\label{eq:dthetadeltaL}
\eeq

The Lagrangian $L$ itself is not diffeomorphism invariant but rather, under a variation which is a diffeomorphism generated
by a vector field $\vec X$, we have
\beq {\bm \delta}_{\vec X} L= {\bm \epsilon} d  {i}_{\vec X}\,L= d{\bm \theta}_{\vec X}\label{eq:delta-L-d-theta}\eeq 
on-shell and all we can deduce from this is that
\beq {\bm \theta}_{\vec X} = {\bm\epsilon} {i}_{\vec X}\,L 
+ {\bm  \WJ}_{\vec X} = {\bm\epsilon}\big( {i}_{\vec X}\,L 
+  \WJ(\vec X\,)\big)\label{chi-def} \eeq
with 
\[\WJ(\vec X) : = \theta(\vec X) - i_{\vec X} L \] 
and $d \WJ({\vec X})=0$. $\WJ(\vec X)$ is referred to as the Noether current in \cite{Wald2}.

Now $ \WJ({\vec X})$ is closed and we 
shall assume that it is $d$-exact on-shell. We shall see below, in equation (\ref{eq:omega-d-phi}), that if this
is not the case there is an obstruction to defining a genuine symplectic
structure on $\widehat {\cal S}$.  So we assume that
\beq \WJ({\vec X}) = d \WQ({\vec X})\,\label{eq:eta-d-psi}\eeq
on-shell and then
\beq {\bm \theta}_{\vec X} = \bepsilon i_{\vec X} L + d {\bm \WQ}_{\vec X}
 = \bepsilon \big(i_{\vec X} L + d \WQ(\vec X)\big).\label{eq:theta-X-d-psi}\eeq
In fact it is argued in \cite{Wald1} that, subject to some mild conditions, $d J(\vec X)=0$ 
implies that $ J({\vec X})$ is $d$-exact but there are some subtleties
in the argument,\footnote{For example it does not apply to
the pre-symplectic density ${\bm \omega}$ which is also $d$-closed (\ref{eq:domega0}).
One of the conditions in \cite{Wald1} for a closed form which depends on some dynamical fields
to be exact is that it  must vanish when the fields vanish.  As long as ${\bm \omega}$
can be put in Darboux form (\textit{e.g} see \cite{Lee+Wald} for the case of Einstein gravity
and \cite{WD} for Einstein gravity with a cosmological constant) it does not
vanish when the fields vanish, because the components are constant in Darboux co-ordinates.} 
so we shall simply assume that $ J({\vec X})$ is $d$-exact.

This then has important consequences for the pre-symplectic density ${\bm \omega}$.
We have
\begin{align*}
{\bm \omega} = \bdelta {\bm\theta}
&=
\big(\hbdelta +\bdelta_{\vec X} \big)\big(\widehat{\bm \theta} + {\bm \theta}_{\vec X}\big)&\\
&=\hbdelta \widehat {\bm \theta} +\bdelta_{\vec X}\, \widehat {\bm \theta} + \hbdelta {\bm \theta}_{\vec X}
+\bdelta_{\vec X}\, {\bm \theta}_{\vec X}& \\
&=\widehat {\bm \omega} +\bdelta_{\vec X}\, \widehat {\bm \theta} + \hbdelta {\bm \theta}_{\vec X},
&
\end{align*}
where $\widehat{\bm\omega}=\hbdelta\widehat{\bm\theta}$ and $\bdelta_{\vec X}\, {\bm \theta}_{\vec X}$
vanishes because ${\bm \epsilon}^2=0$.
Now
\begin{align*}
\bdelta_{\vec X}\, \widehat {\bm \theta} + \hbdelta {\bm \theta}_{\vec X}
&={\bm\epsilon}{\cal L}_{\vec X}\, \widehat {\bm \theta} 
+\hbdelta \big({\bm \epsilon} i_{\vec X} L + \bepsilon  \WJ(\vec X\,)\big) & \hbox{using (\ref{chi-def})}\\
&= {\bm \epsilon} \big((d i_{\vec X}\, \widehat{\bm\theta}+ i_{\vec X}\, d\widehat{\bm\theta}\,)
 -  i_{\vec X}\,\hbdelta L - \hbdelta  d \WQ(\vec X\,)\big)&\\ 
&={\bm\epsilon} \big( d i_{\vec X}\,\widehat{\bm\theta} - \hbdelta d \WQ(\vec X\,)\big)&
\hbox{using (\ref{eq:dthetadeltaL})}\\
& :=  \bm \epsilon d \bm\phi(\vec X)
\end{align*}
where
\beq \bm\phi(\vec X) = i_{\vec X} \widehat {\bm \theta} - \hbdelta Q(\vec X) \mod d. \eeq
Since
\[\bdelta_{\vec X}\, \widehat {\bm \theta} + \hbdelta {\bm \theta}_{\vec X} = {\bm \omega}(\bdelta_{\vec X}F^I,\hbdelta F^I) \]
this relates to the Hamiltonian flow \cite{Wald2} --- if there is a Hamiltonian ${\cal H}[\vec X]$ on phase space 
that generates the evolution corresponding to the flow arising from $\vec X$ (not necessarily time-like) then 
\[ {\bm \omega}(\bdelta_{\vec X}F^I,\hbdelta F^I)=-\hbdelta {\bm h}_{\vec X}
={\bm\epsilon}\hbdelta h(\vec X)\]
with
\[\hbdelta h(\vec X)= d \bm \phi(\vec X)
=d\big\{i_{\vec X}\widehat {\bm \theta}- \hbdelta Q(\vec X)\big\}\]
the variation of the corresponding Hamiltonian density. In that case we can define
a Hamiltonian ${\cal H}[\vec X]$ which satisfies
\[ \hbdelta {\cal H}[\vec X] = \int_\Sigma \hbdelta h(\vec X)
= \int_{\partial \Sigma }\big(i_{\vec X}\widehat{\bm \theta} -\hbdelta  \WQ (X) \big).
\]
If $\vec X$ does not vanish on the boundary and
\[ \int_{\partial \Sigma} i_{\vec X}\widehat{\bm \theta}\ne 0\]
then a Hamiltonian corresponding to the flow generated by $\vec X$
will only exist if 
$\int_{\partial \Sigma} i_{\vec X}\widehat{\bm \theta}$ is $\bdelta$-exact.
The existence of such a Hamiltonian requires this integrability condition
which is decided by the specific theory in question, \cite{Wald2}.

We can see that  $h(\vec X)$ and $J(\vec X)$ are well defined in cohomology.  Under the change
$L\rightarrow  L + d\alpha$ and $\bm\theta \rightarrow  \bm\theta + \bdelta \alpha + d \bm \psi$,
\[ \bm J_{\vec X} = \bm\theta_{\vec X} - \bm\epsilon i_{\vec X} L 
\ \rightarrow \ 
\bm J_{\vec X} +d(\bm \epsilon i_{\vec X}\alpha + \bm \psi_{\vec X})=
\bm \epsilon \big\{ J(\vec X) + d\big( i_{\vec X}\alpha + \psi(\vec X)\big)\big\} \]
is unchanged if we choose $\psi(\vec X) = -i_{\vec X} \alpha \mod d$. This together with
\[ \widehat {\bm \theta}  \rightarrow   \widehat {\bm \theta} +\hbdelta \alpha + d \widehat {\bm \psi}, \]
then shows that 
\bean \bm \epsilon \hbdelta h(\vec X) & \rightarrow &  \bm \epsilon \hbdelta h(\vec X) -\hbdelta \bm \epsilon d\big( i_{\vec X}\alpha + \psi(\vec X)\big) - \bm \epsilon d i_{\vec X} (\hbdelta \alpha + d \widehat {\bm \psi} )\\
&=&  
\hbdelta \bm h_{\vec X}  + \bm \epsilon \big\{\hbdelta d \psi(\vec X) - d i_{\vec X} d \widehat {\bm \psi} \,\big\} \eean
is unchanged provided we choose $\widehat {\bm \psi}$ such that 
\[i_{\vec X} d \widehat {\bm \psi} = \hbdelta \psi(\vec X) =- i_{\vec X} \hbdelta \alpha \mod d.\]

In any case, whether or not a Hamiltonian exists, we have, assuming $\WJ=d\WQ$ on-shell,
\beq
\bdelta_{\vec X}\, \widehat {\bm \theta} + \hbdelta {\bm \theta}_{\vec X}
=d{\bm \phi}_{\vec X}
\label{eq:delta-theta-dphi}\eeq
with
\beq{\bm \phi}_{\vec X} :=   {\bm \epsilon} i_{\vec X}\,\widehat{\bm \theta}+\hbdelta {\bm \WQ}_{\vec X} \mod d
 \eeq
and
\[{\bm \epsilon} i_{\vec X}\,\widehat{\bm \theta}+\hbdelta {\bm \WQ}_{\vec X} ={\bm \epsilon} \big(i_{\vec X}\,\widehat{\bm \theta}-\hbdelta \WQ(\vec X\,)\big).\]

This results in  the important conclusion that, for a diffeomorphism $\vec X$,
\beq {\bm \omega} = \widehat{\bm \omega} + d{\bm \phi}_{\vec X}\label{eq:omega-d-phi}\eeq
and this guarantees that 
\[\widehat{\bm \omega} = \hbdelta \widehat{\bm \theta} + \hbdelta   {\bm \phi}_{\vec X}\]
is a bona fide symplectic density on $T^*\widehat{\cal S}$ when $\Sigma$ is compact without boundary,
because it pulls back to ${\bm\omega}\mod d$ under the projection from ${\cal S}$ to $\widehat{\cal S}$
as demanded in \cite{C-W}.
If $\Sigma$ is compact without boundary
it follows that $\widehat{\bm \Omega}$ on $\widehat {\cal S}$ pulls back to the pre-symplectic form ${\bm \Omega}$ on $T^*{\cal S}$.
If $\Sigma$ has a boundary $\partial \Sigma$ then we must restrict
the diffeomorphisms to allow only those for which
the vector fields generating them fall off fast enough at the 
boundary so that surface terms vanish.

The essence of the above formulae is summarised in the following on-shell diagram

\beq \kern -50pt\xymatrix@=50pt@M=10pt{
  L \ar[r]^{\kern -40pt\hbdelta} \ar[rd]^{ {{\bm \epsilon} i}_{\vec X}}
  & \boxed{
\begin{array}{cc}
\widehat {\bm \delta} L &= d \widehat {\bm \theta}\\
{\bm \epsilon} d i_{\vec X}L &= d{\bm \theta}_{\vec X}\, \end{array}
} \ar[rd]^{\bepsilon i_{\vec X}}
\ar[r]^{\kern 20pt \hbdelta} & 0 \\
& \ar[u]_d  \boxed{    
\begin{array} {c}
\widehat {\bm \theta} \\
{\bm \theta}_{\vec X}\,={\bm \epsilon} i_{\vec X}\, L + d{\bm \WQ}_{\vec X}
\end{array}
}
  \ar[r]^{\kern -25pt\hbdelta} \ar[rd]^{{ {{\bm \epsilon} i}_{\vec X}}} & 
\boxed{ 
\begin{array}{cc}
\widehat {\bm \omega} = \hbdelta \widehat {\bm \theta} \\
\hbdelta{\bm \theta}_{\vec X}\, + \bdelta_{\vec X}\,\widehat  {\bm \theta}
=  d{\bm\phi}_{\vec X}
\end{array}
}  \ar[u]_d  \\
 & \ar[u]_d  \boxed{  {\bm \WQ}_{\vec X} }
\ar[r]^{\kern -45pt \hbdelta} & 
 \ar[u]_d 
\boxed{
{\bm \phi}_{\vec X}
   ={\widehat {\bm \delta}}{\bm \WQ}_{\vec X}  + {{{\bm \epsilon} i}_{\vec X}}\,\widehat {\bm \theta}\,
  }\,.\ 
} \label{eq:xy-ix-delta-d}\eeq
Integrating the upper member of the middle row over $\Sigma$ gives
\[\widehat \Theta = \int_{\Sigma} \widehat {\bm \theta}
\quad \mathop{\longrightarrow}^{\hbdelta} \quad \widehat \Omega =  
\int_{\Sigma} \widehat {\bm \omega}.\] 

We have here a double complex \cite{Bott+Tu} whose general structure is
\[ \xymatrix@=50pt@M=10pt{
W^{0,n+1}\ar[r]^\hbdelta  \ar[rd]^{ {{\bm \epsilon} i}_{\vec X}}   &  W^{1,n+1}  \ar[r]^\hbdelta \ar[rd]^{ {{\bm \epsilon} i}_{\vec Y}}  &  W^{2,n+1}  & \kern -120pt \cdots\\
W^{0,n} \ar[u]_d \ar[r]^\hbdelta \ar[rd]^ {{\bm \epsilon} i_{\vec X}} & 
\ar[u]_d \ar[r]^\hbdelta  W^{1,n}  \ar[rd]^{{\bm \epsilon} i_{\vec Y}}  & \ar[u]_d  W^{2,n}
& \kern -120pt  \cdots\\
\vdots \ar[r]^\hbdelta \ar[rd]^{{\bm \epsilon} i_{\vec X}} \ar[u]_{d} &\vdots \ar[r]^\hbdelta 
\ar[rd]^{{\bm \epsilon} i_{\vec Y}}  \ar[u]_d & \vdots  \ar[u]_d & \kern -120pt  \cdots\\
W^{0,0} \ar[u]_d\ar[r]^\hbdelta  & \ar[u]_d  W^{1,0}\ar[r]^\hbdelta    & \ar[u]_d W^{2,0}
& \kern -120pt  \cdots\\
} \] 
In principle subsequent diffeomorphisms could be generated by different 
vector fields $\vec X$ and $\vec Y$, though the analysis above assumed
$\vec X = \vec Y$ in order to understand how a single diffeomorphism affected ${\bm \omega}$ in (\ref{eq:omega-d-phi}).

Note that ${\bm \epsilon} i_{\vec X}$ preserves the total degree $p+q$ of the forms
on $W^r$.
\subsection{Killing symmetries and conserved charges}

The formalism really comes into its own for discussing symmetries.
If the classical action has symmetries that are broken at the quantum level
there are anomalies and the double complex sketched in (\ref{eq:xy-ix-delta-d}) is the natural framework for 
analysing the Stora-Zumino descent 
equations. These can include
gravitational anomalies associated with local Lorentz invariance and diffeomorphism anomalies, which are equivalent \cite{Gaume-Ginsparg}, but we shall not analyse anomalies here. 

Consider (\ref{eq:delta-theta-dphi}) when $\vec X=\vec K$
\beq
\bdelta_{\vec K} \widehat {\bm \theta} + \hbdelta {\bm \theta}_{\vec K}
= \hbdelta {\bm \theta}_{\vec K}=d{\bm \phi}_{\vec K}
\label{eq:delta-theta-dphi-K}\eeq
and, for $\vec K$ Killing,
\[ \bepsilon {\cal L}_{\vec K} L = \bepsilon d i_{\vec K}L= d {\bm \theta}_{\vec K}=0\]
so ${\bm \theta}_{\vec K}$ is $d$-closed.  
In fact since ${\bm \theta}_{\vec K}$ is linear in ${\cal L}_{\vec K} F^I$ and 
${\cal L}_{\vec K} \partial^k F^I$ we expect that ${\bm \theta}_{\vec K}=0$ when 
${\vec K}$ is Killing.\footnote{For gauge theories this might not be strictly true. For electromagnetism, for example, $L=-\frac{1}{2}F\wedge *F$ and ${\bm \theta}_{\vec K} = - \bm \epsilon ({\cal L}_{\vec K} A) \wedge *F$. While ${\cal L}_{\vec K}*F=0$ by assumption
${\cal L}_{\vec K}A$ might not be. But on-shell ${\bm \theta}_{\vec K }=- \bm \epsilon (i_{\vec K} F) \wedge *F \mod d$ and so is invariant$\mod d$  under a gauge transformation.}  For the same reason $\bdelta_{\vec K} \widehat {\bm\theta}=0$
and 
\beq d {\bm \phi}_{\vec K}=0\label{eq:dphi-0}\eeq
with
\[{\bm \phi}_{\vec K} 
={\bm \epsilon} i_{\vec K}\,\widehat{\bm \theta}+\hbdelta {\bm \WQ}_{\vec K}
={\bm \epsilon} \big(i_{\vec K}\,\widehat{\bm \theta}-\hbdelta \WQ(\vec K)\big).\]

Two important conclusions immediately follow:
\begin{itemize}
\item From (\ref{eq:omega-d-phi}) and (\ref{eq:dphi-0})
\[{\bm\omega}=\widehat{\bm \omega}\]
for a Killing symmetry.

\item If we can foliate $\Sigma$ into hypersurfaces $\sigma_r$ (\textit{e.g.} $r$ could be a radial co-ordinate) then we can integrate over a piece of $\Sigma$
which is a thick shell $\Sigma_{[r,r']}$  between $r$ and $r'$ and 
\[ \int_{ \Sigma_{[r,r']}} d {\bm \phi}_{\vec K}  = \int_{\sigma_{r'}} {\bm \phi}_{\vec K}
- \int_{\sigma_r} {\bm \phi}_{\vec K}= 0\]
implies that
\[{\bm \Phi}[\vec K\,]:= \int_{\sigma_r}{\bm \phi}({\vec K}) =\int_{\sigma_r} \big(i_{\vec K}\widehat {\bm \theta} - \hbdelta \WQ(\vec K\,)\big),\]
a 1-form on $T^*{\cal S}$,
is independent of $r$.
For example  $\sigma_r$ might be an $(n-1)$ sphere
and it can be convenient to evaluate ${\bf \Phi}$ at $r\rightarrow \infty$
but the formalism here shows that this is not essential. Any value of $r$
can be used in principle, though in practice it is usually easier to do the
integrals at $r\rightarrow \infty$. It is not even necessary to use a round sphere.
\end{itemize}

\noindent Recalling the discussion of the Hamiltonian there may be an obstruction to obtaining a genuine charge from
${\bm \Phi}[\vec K\,]$, it is a 1-form on $T^*{\cal S}$ and does not yet yield
a charge.  A genuine charge emerges from this construction only if
$i_{\vec K}\widehat {\bm \theta}$ is $\hbdelta$-exact.  If this is this case,
and only if this is the case, we can write
\beq i_{\vec K}\widehat {\bm \theta} = \hbdelta \mu(\vec K) \label{eq:iKtheta-dmu}\eeq
and define
\beq \rho(\vec K) = \mu(\vec K)-Q(\vec K), 
\eeq
with $\rho(\vec K)$ an $(n-1)$-form on ${\cal M}$ satisfying
\[ \hbdelta \rho(\vec K) =\bm \phi(\vec K), \qquad \hbdelta h(\vec K) = \hbdelta d \rho(\vec K).\]
When $\vec K$ is Killing $d{\bm \phi}(\vec K)=0$
and we then have an invariant
\[\hbdelta {\cal H}[\vec K] = \int_{\partial \Sigma}\hbdelta {\rho(\vec K)}=0.\]
Again if $\partial \Sigma = \sigma_r \cup \sigma_{r'}$
\[ \hbdelta{\cal Q}:=\int_{\sigma_r}\hbdelta {\rho(\vec K)} = \int_{\sigma_{r'}}\hbdelta {\rho(\vec K)}\]
is independent of which copy of $\sigma$ it is evaluated on.
We can associate a Noether charge\footnote{In general this is not the same as Wald's Noether charge associated with the entropy in \cite{Wald2}.}
 \beq {\cal Q}[\vec K\,] = \int_{\sigma} \rho(\vec K\,)\eeq
with the symmetry ${\vec K}$, for which
\[\hbdelta {\cal Q}[\vec K] ={\bm \Phi}[\vec K].\]
For example for a stationary space-time, with Killing vector $\vec K=\frac{\partial}{\partial t}$, $ {\cal Q}[\vec K]$ is a mass while for an axially symmetric space-time, with Killing vector $\vec K = \frac{\partial}{\partial \varphi}$, ${\cal Q}[\vec K]$ is the angular momentum associated with the space-time.
   
In summary, given a solution of the equations of motion with Killing vector ${\vec K}$, we define the $(n-1)$-form $\WQ(\vec K)$ mod $d$ by 
\[ d \WQ(\vec K) =\theta(\vec K)- i_{\vec K}L.\]
Then, if and only if $i_{\vec K}{\bm \theta} = \hbdelta \mu(\vec K)$ is $\bdelta$-exact, we have
\[ \hbdelta  {\cal Q}[\vec K] = \hbdelta \int_{\sigma_r} \rho(\vec K) = 
\int_{\sigma_r} \big( i_{\vec K}\widehat{\bm \theta}-\hbdelta \WQ(\vec K) \big).\]

In analogy with (\ref{eq:Noether-current}) we define a  2-form to
be the Hodge dual of $\rho(\vec K\,)$,
\beq \rho(\vec K\,)=*\NJ(\vec K\,),\label{eq:Noether-2-form}\eeq
and
\[  {\cal Q}[\vec K\,] = \int_{\sigma} *\NJ(\vec K\,)\]
with $d*\NJ=0$ on-shell.
For a space-time symmetry generated by a Killing vector $\vec K$ the analogue
of a Noether current is a Noether 2-form.

We have the sequence
\[ \xymatrix@=50pt@M=10pt{
 {\cal Q} \ar[r]_\hbdelta & {\bm \Phi}\ar[r]_\hbdelta & 0\, .} \]

\section{Example: Einstein gravity and the\\ Schwarzschild geometry}

As an example of these ideas example consider Einstein gravity in four dimensions.
The Lagrangian is
\[L = \frac{1}{16\pi} R_{a b} \wedge * e^{a b}\]
where $e^a$, $a=0,1,2,3$ are orthonormal 1-forms,
\[ R_{a b} = d \omega_{a b} + \omega_{a c} \wedge \omega^c{}_b,\]
are the curvature 2-forms for the associated connection 1-forms $\omega_{a b}$
and $*$ is the Hodge star. The connection 1-forms are determined by the
torsion free condition
\[ D e^a = d e^a + \omega^a{}_b \wedge e^b=0\]
and orthonormal indices are lowered with $ \eta_{a b}=\hbox{diag}(-1,0,0,0)$.

Under a variation of the 1-forms, $e^a \rightarrow e^a + \delta e^a$, 
\[ \delta R_{a b} = D (\delta \omega_{a b})\]
and 
\[\delta L = \frac{1}{16\pi} \big\{\delta e^c\wedge E_c
+ D(\delta \omega_{a b}) \wedge * e^{a b}\big\}
=\frac{1}{16\pi} \big\{(\delta e^c \wedge E_c
+  d\big((\delta \omega_{a b}) \wedge * e^{a b}\big)\big\}\]
where 
Einstein's equations are
\[  E^c = R_{a b} \wedge * e^{a b c}=0.\]
From this we get\footnote{There is a subtlety here, not all $\delta e^a$ correspond to genuine variations in the metric some are just local tangent space rotations. Expanding $\delta e^a = \Delta^a{}_b e^b$ only the
symmetric part of $\Delta_{a b}$ can give genuine metric variations (and some of these are just diffeomorphisms), the anti-symmetric part of $\Delta_{a b}$ is a tangent space rotation (Lorentz transformation). We ignore this problem here and just choose a gauge in which $\Delta_{a b}$ is symmetric, but this is not necessary.  
This relates to the fact that (\ref{eq:deltaL}) only defines ${\bm \theta}$ mod $d$, in general the anti-symmetric part of $\Delta_{a b}$ can be eliminated by adding a $d$-exact form to ${\bm \theta}$. Full details are given in \cite{WD}.}
\beq \theta(\delta e^a) = (\delta \omega_{a b}) \wedge * e^{a b}.\label{E-theta}\eeq
 
For a diffeomorphism, $\delta e^a = {\cal L}_{\vec X}\,e^a$,
\beq \theta({\cal L}_{\vec X}\,e^a) =  
\frac{1}{16\pi}({\cal L}_{\vec X}\,\omega_{a b}) \wedge * e^{a b} 
=  -\frac{1}{16\pi} d*d X \label{eq:theta-xi}\eeq
on-shell.  Also
\bea{\cal L}_{\vec X}\, \theta(\delta e^a)
&=& \frac{1}{16\pi}\big\{d i_{\vec X}\,\big(\delta \omega_{a b} \wedge * e^{a b}\big)
+i_{\vec X}\,d \big(\delta \omega_{a b} \wedge * e^{a b}\big)\big\}
\nonumber\\
&=& \frac{1}{16\pi}\big\{d i_{\vec X}\,\big(\delta \omega_{a b} \wedge * e^{a b}\big)
+i_{\vec X}\,\big( D(\delta \omega_{a b}) \wedge * e^{a b}\big)\big\}\nonumber\\
&=& \frac{1}{16\pi}\big\{d i_{\vec X}\,\big(\delta \omega_{a b} \wedge * e^{a b}\big)
+i_{\vec X}\,\big( (\delta R_{a b}) \wedge * e^{a b}\big)\big\}.\label{eq:ell-X}
\eea
By assumption $\delta e^a$ satisfies the linearised equations of motion so
\[ \delta R_{a b} \wedge * e^{a b c} = 0 \quad \Rightarrow \qquad 
\delta R_{a b} \wedge * e^{a b} = 0.\]
Combining (\ref{eq:theta-xi}) and (\ref{eq:ell-X}) we conclude that the pre-symplectic density satisfies
\bea \omega({\cal L}_{\vec X}\, e^a, \delta e^a,) 
&=&{\cal L}_{\vec X}\, \theta(\delta e^a)
- \delta \theta({\cal L}_{\vec X}\,e^a) \nonumber\\
&=&
(\delta \omega_{a b}) \wedge  ({\cal L}_{\vec X} * e^{ a b})
-  ({\cal L}_{\vec X}\omega_{a b}) \wedge (\delta * e^{ a b}) \nonumber \\
&=& \frac{1}{16\pi} d\big\{
i_{\vec X}\,\big(\delta \omega_{a b} \wedge * e^{a b}) + \delta (* d X)\big\}
\eea
and we have obtained 
\beq \phi(\vec X\,) = \frac{1}{16\pi} \big\{
i_{\vec X
}\,\big(\delta \omega_{a b} \wedge * e^{a b}) + \delta (* d X)\big\}\label{eq:phi-def}\eeq
for Einstein gravity.
When $\vec X=\vec K$ is Killing both $\theta({\cal L}_{\vec K}\, e^a)$ and the symplectic form $\omega(\delta e^a,{\cal L}_{\vec K}\, e^a)$ vanish and $*d K$ is the Komar 2-form.

\subsubsection{Derivation of Noether mass for the Schwarzschild solution}

To illustrate the above ideas we consider the Schwarzschild metric
\[d s^2 = -\left(1 -\frac{2 m}{r} \right) d t^2 +  
\left(1 -\frac{2 m}{r} \right)^{-1} d r^2 + r^2 (d\vartheta^2 + \sin^2 \vartheta \,d \varphi^2).\] 
We choose orthonormal 1-forms
\[\begin{array}{ll}
e^0  = \sqrt{1-\frac{2 m}{r}}\, d t, \kern 20pt & e^2 = r d \vartheta, \\
e^1  = \frac{d r}{\sqrt{1-\frac{2 m}{r}}},  & e^2 = r\sin\vartheta d\varphi,
\end{array}\]
giving connection 1-forms
\beq \begin{array}{lll}
\omega_{0 1} & = -\frac{m}{r^2} d t, &\omega_{0 2}= \omega_{0 3}=0,\\
\omega_{1 2} &= -\sqrt{1-\frac{2 m}{r}}\, d\vartheta,
\kern 20pt &  \omega_{1 3} = -\sqrt{1-\frac{2 m}{r}}\, \sin\vartheta \,d\varphi, \\
\omega_{2 3}& = -\cos\vartheta\, d\varphi. &
\end{array}\label{eq:Schwarz-connection} \eeq
We shall calculate the mass associated with the Killing vector
\[ \vec K = \frac{\partial}{\partial t}\]
and its metric dual 1-form
\[ K=-\left(1-\frac{2 m}{r}\right) d t\]
(with signature $(-,+,+,+)$).
We have
\beq d K = \frac{2 m}{r^2}  e^{0 1}\label{eq:dK}\eeq
and 
\beq * d K = \frac{2 m}{r^2} e^{2 3} = 2 m \sin\vartheta\, d\vartheta \wedge d\varphi \label{eq:star-d-K}\eeq
(we use conventions with $1=e^{0123}$ and $*e^{0 1} = e^{2 3}$).
Note that $d*dK=0$ so $\theta(\vec K\,)=0$.

Now suppose the metric variation $\delta e^a$ is induced by varying the parameter $m$.  For this variation
\[ \delta e^0 = -\frac{\delta m}{r- 2m} e^0,
\quad \delta e^1 = \frac{\delta m}{r- 2m} e^1,
\quad \delta e^2 = \delta e^3=0,\]
with $\delta m$ constant,
while (\ref{eq:Schwarz-connection}) gives
\bea
\delta \omega_{0 1} &= -\frac{\delta m}{r^2} \frac{e^0}{\sqrt{1-\frac{2 m}{r}}},\nonumber  \\
 \delta \omega_{1 2} &= \frac{\delta m}{r^2} \frac{e^2}{\sqrt{1-\frac{2 m}{r}}},\label{eq:Schwarz-delta-omega}\\
 \delta \omega_{1 3} &= \frac{\delta m}{r^2} \frac{e^3}{\sqrt{1-\frac{2 m}{r}}}.\nonumber
\eea
The variations (\ref{eq:Schwarz-delta-omega}) produce
\[ i_{\vec K} (\delta \omega_{a b}\wedge *e^{a b}) 
=\frac{2 \delta m}{r^2} e^{23}.\]
Also from  (\ref{eq:star-d-K}) 
\[\delta(*d K) = \frac{2 \delta m}{r^2} e^{2 3} \] 
and these combine in (\ref{eq:phi-def}) to give
\[ \phi(\vec K\,) = \frac{1}{4\pi}\frac{\delta m}{r^2} e^{23}
=\frac{\delta m}{4\pi}  \sin\vartheta\, d\vartheta\, d\varphi. \] 
Now we choose constant time-slices $\Sigma$ with $2m < r < \infty$, $0\le\vartheta \le \pi$  and $0\le \varphi < 2\pi$.  $\sigma_r$ are 2-spheres of radius $r$, and
\[\delta {\cal Q}(\vec K\,)= \int_{\sigma_r}\phi(\vec K\,) = \frac{\delta m}{4\pi} \int_0^\pi \int_0^{2\pi}\sin\vartheta\, d\vartheta\, d\varphi = \delta m\]
so
\beq  {\cal Q}=\frac{1}{4\pi} \int_{S^2} m  \sin\vartheta d \vartheta d \varphi,
= m.\label{eq:Gauss}\eeq
The parameter $m$ in the Schwarzschild metric is indeed the mass,
as expected.

Note that the final answer is independent of $r$, it is not necessary to take $r\rightarrow \infty$ in order
to calculate the mass.  Indeed in this example
\[ \NJ = \frac{m}{4 \pi r^2} e^{0 1}\]
and (\ref{eq:Gauss}) is exactly analogous to Gauss' law in electrostatics.
This similarity between the Maxwell 2-form field strength $F$ and $\NJ=dK$ for a Killing vector
$\vec K$ was pointed out in \cite{Simon}.

\section{The relation between the Noether current and the Noether 2-form}

In this section we expand further on the relation between the Noether current $j$ and the Noether 2-form $J$ for space-time symmetries. 
Under a variation of the fields
\[ \delta L = d \theta\]
on-shell.
The conventional Noether current is obtained from a symmetry generator ${\cal T}_ \NQ$
under which the Lagrangian is invariant on-shell,
\[ {\cal T}_ \NQ L=d\theta=0\]
and
\[\theta=*j. \]
For a Killing symmetry generated by a vector field $\vec K$
\[{\cal L}_{\vec K}L=d i_{\vec K} L\]
and
\[ d \{ \theta({\vec K})- i_{\vec K} L \}=0.\] 
If $\theta({\vec K}) -i_{\vec K} L$ is $d$-exact we define $\WQ(\vec K)\mod d$ via
\[ d \WQ(\vec K) = \theta(\vec K)-i_{\vec K}L.\] 
Under a genuine variation of the dynamical fields, which is not a diffeomorphism,
\[ \widehat \delta L = d \widehat\theta.\]
and, if $i_{\vec K} \widehat \theta$ is $\widehat\delta$-exact,
\[ i_{\vec K} \widehat \theta = \widehat\delta \mu(\vec K).\]
The 2-form $\NJ(\vec K)$ (that is a 2-form on $T^*{\cal M}$) is then defined via
\[\widehat \delta *\NJ =  \widehat \delta\mu-\widehat \delta\WQ . \]

The definition of $\NJ$ is very different to the standard approach to the Noether current for space-time symmetries associated with a Killing vector.
The standard approach assumes that the Lagrangian can be decomposed into a \lq\lq geometrical'' term and a \lq\lq matter'' term, \textit{e.g} for Einstein gravity
\[ L= -\frac {1}{16\pi} R_{a b}\wedge *e^{a b} + L_{Matter}.\]
If the metric is not dynamical only $L_{Matter}$ is considered.
The energy-momentum tensor is defined by varying the metric in $L_{Matter}$, even if it is not dynamical.
In terms of orthonormal 1-forms
\[ \delta_{e^a} (L_{Matter}) = \delta e^a \wedge \tau_a\]
where
\[ \tau_a = T_{a b} * e^b.\]
Conservation of energy-momentum can be expressed as
\[ D \tau ^a =0 \qquad \Leftrightarrow \qquad D_a T^a{}_b=0\]
and we assume this is true on-shell. 
Then for a diffeomorphism $\delta e^a =  {\cal L}_{\vec X}e^a$ and
\bean
 \delta_{\vec X} L_{Matter} &=& \big(d i_{\vec X} e^a + i_{\vec X} d e^a\big)\wedge \tau_a \\
&=& \big(d X^a + \omega^a{}_b  X^b - (i_{\vec X}\omega^a{}_b) e^b  \big)\wedge \tau_a\\ 
&=& (D X^a)\wedge \tau_a\\
&=&  d( X^a \tau_a), 
\eean
where we have used $e^b\wedge\tau^a=T^{a b}*1$ and $\omega_{a b} = - \omega_{b a}$.
If $\vec X = \vec K$ is Killing then 
\[ \delta_{\vec K} L_{Matter}=  d (K^a \tau_a) = 0\]
on-shell and the Noether current
\[ *j= K^a \tau_a \]
is conserved, $d*j=0$ (in components $j_b = K^a T_{a b}$).

This is clearly on a different footing to conservation of the 2-form, 
\hbox{$d*\NJ=0$}.
The d-cohomology classes of $*j$ and $*\NJ$ are completely different --- they carry different
geometrical information. 

\section{Discussion}

Wald and collaborator's description of diffeomorphic invariant theories 
(the generalisation of Witten and Crnkovi\'c description of
Yang-Mills theories and general relativity) and the construction of symplectic structures and Noether charges fits naturally into a double complex structure,
summarized in the commutative diagram (\ref{eq:xy-omega-delta-theta}).
This mathematical structure is also relevant to quantum anomalies. The field variations ${\bm \delta}$ include symmetry transformations of the classical action and these can generate new terms if there is a quantum anomaly. The double complex not only gives a covariant description of phase space and the symplectic structure as will as classical invariants it also describes the cohomology of quantum anomalies.

The explicit example of the Schwarzschild metric shows how the time-like Killing
vector generates the Noether charge associated with the mass. This can
be calculated exactly on any sphere surrounding the event horizon, it is not necessary to perform the calculation at $r\rightarrow \infty$ (the only role the asymptotic regime plays is to furnishes the correct normalisation for the Killing vector, which is chosen to give $\frac{\partial}{\partial t}$ unit length
only at $r\rightarrow \infty$). This is not in itself a new result --- it was shown that the Noether charge correctly reproduces the ADM mass for asymptotically flat space-times in \cite{Iyer+Wald} --- but the same formalism also correctly reproduces the Brown-York mass and the Bondi mass as well as the Henneaux-Teitelboim mass for a rotating black hole in asymptotically anti-de Sitter space-time, the details of these will be published elsewhere \cite{WD}.

\end{document}